# Preventing Distinguishability between Multiplication and Squaring Operations

Alkistis Aikaterini Sigourou[1], Zoya Dyka[1,2], Peter Langendoerfer[2] and Ievgen Kabin[1]
[1] *IHP – Leibniz-Institut für innovative Mikroelektronik,* Frankfurt (Oder), Germany
[2] *BTU Cottbus-Senftenberg,* Cottbus, Germany
{sigourou, dyka, kabin}@ihp-microelectronics.com

*Abstract*— Scalar multiplication *kP* is a critical operation in Elliptic Curve Cryptosystems (ECC), often targeted by Side-Channel Analysis (SCA). Despite strategies based on atomic patterns to enhance security, the binary *kP* algorithms remain susceptible to simple SCA due to energy consumption variations in field multipliers during passing two different or two identical operands. This vulnerability arises independent of the multiplication method used. We implemented and analysed two mitigation techniques: one involving data redirection and another focusing on bus reloading.

## I. Introduction

Elliptic Curve Cryptosystems (ECC) are critical for securing communication, utilizing the cryptographic protocols for digital signature, authentication and secret key exchange. Central to these protocols is the Elliptic Curve (EC) point-scalar multiplication, which is denoted as *kP* operation and is susceptible to attacks aiming to reveal the secret value of the scalar *k*. To mitigate risks, Hardware Security Modules (HSM) have to be resistant to various physical attacks, including side-channel attacks (SCA), when attackers have physical access to the HSM and can measure and analyse power or electromagnetic traces of *kP* executions. Notably, horizontal SCA attacks, which examine a single *kP* execution trace, pose significant threats. Binary *kP* algorithms are mostly implemented in hardware and process the binary value of the scalar *k* bit-by-bit. Attackers by applying visual inspection, different statistical, machine learning, or other methods, try to distinguish and classify parts of the trace related to each bit of *k*, separating them into two subsets for '0' and '1' using differences in the slot shapes. Regularity and atomicity principles are the basis for countermeasures against simple SCA attacks. Regular algorithms such as the Montgomery ladder based on [1] or the double-and-add-always *kP* algorithm [2] ensure that each bit of the scalar *k* is processed identically, making it difficult to differentiate the shape of '0'-slots and of '1'-slots. Atomic block patterns, for example [3], [4], [5], [6], represent each slot as a set of atomic blocks (atoms), whereby each atom consists of the same sequences of operations. That makes all atoms identical from SCA point of view. Hardware implementations of regular and atomic pattern binary algorithms are vulnerable to horizontal address-bit SCA attacks, with no known effective algorithmic countermeasures available [7]. A new marker exploiting the distinguishability of multiplication and squaring of finite field elements is published in [8], [9].

In this paper, we investigate the example of our implementation of the *kP* operation using Longa's atomic patterns [4] to determine whether dummy addressing and data flow redirection can enhance the resistance of hardware implementations of *kP* operations.

## II. Implementation Details

We implemented the binary double-and-add left-to-right scalar multiplication algorithm for the NIST P-256 elliptic curve using Longa's atomic patterns [4]. Each of Longa's atomic blocks follows the MNAMNAA sequence of field operations, with M = multiplication, N = negation, and A = addition. Processing a '0' bit-value of *k* requires 4 such atoms, while processing a '1' bit-value of *k* consists of 10 atoms.

Our *kP* design consists of functional blocks for addition, subtraction, and multiplication in the prime finite field *GF(p)*, utilizing 256-bit registers for operand storage. A Controller organizes the sequence of field operations and data transfers between blocks via a multiplexer, ensuring only one block or register can transfer its output value to the bus in each clock cycle while many blocks can read from it. Using Cadence's SimVision v. 15.20-s053, we synthesised the design for the IHP 250 nm cell library SGB25V for a clock cycle period of 30 ns and simulated a power trace of the design during a single *kP* execution. Implementation details, on the scalar *k* and EC point *P* and the identified vulnerability are elaborated in [9], whereas here we primarily highlight the vulnerability and concentrate on the implemented countermeasures.

A comparison of atomic blocks shows that even when blocks appear structurally similar, the actual register usage and operations differ. In particular, the power shapes of regular multiplication and squaring differ due to data-bit and address-bit reasons. This difference is short, only 1 clock cycle, but significant, and can be exploited for successfully revealing the scalar *k*, which is the secret and target of the attack. Fig. 1 illustrates the power profile comparison between field multiplication and field squaring operations in the design.

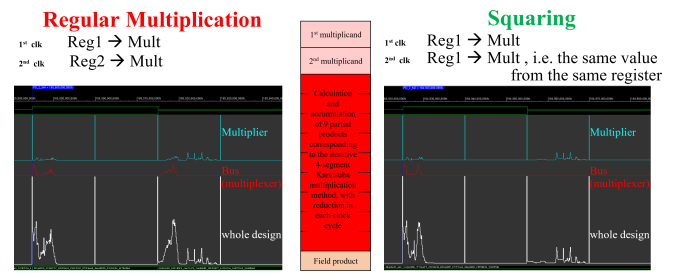

Fig. 1. The distinguishability in the power trace of regular multiplication and squaring of finite field elements; includes field multiplication: schematic representation

In the first clock cycle of both operations, the Controller addresses the register for the first multiplicand and sends it to the field multiplier through a multiplexer. During multiplication, the two multiplicands are distinct, requiring addressing different registers, thus yielding new values. In the second clock cycle of multiplication, a new register is accessed, resulting in energy consumption due to activity in the multiplexer, which is indicated by the red line in the

Regular Multiplication trace. In contrast, during field squaring, the second multiplicand is the same as the first one, keeping both the address and value constant in the second clock cycle. This leads to no changes in the multiplexer, resulting in no energy usage as indicated by the flat red line in the Squaring trace. The multiplexer, a substantial component of the *kP* architecture, significantly impacts overall energy consumption, highlighting a distinct power difference between squaring and regular multiplication in the second clock cycle, attributed to data-bit and address-bit effects.

To prevent the distinguishability described above, designers can employ several strategies.

One approach is to insert a dummy operation between the transfers of the first and second multiplicand. For example, the multiplier can obtain a dummy 2$^{nd}$ operand. Alternatively, any inactive block of the design can be addressed to obtain a dummy operand if the block is not storing an intermediate value important for the correct calculation. Moreover, any block, for example, the field multiplier, can be instructed to write its output to the bus in a seemingly random manner, making the sequence unpredictable to attackers. This technique requires one additional clock cycle for each multiplication or squaring operation. Fig. 2 shows the power shapes of the first 3 clock cycles for the design implementing this approach.

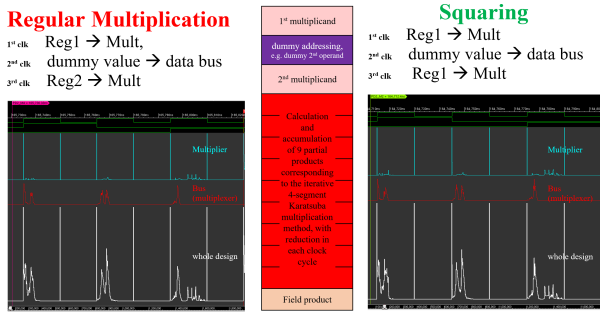

Fig. 2. Mitigating multiplication–squaring distinguishability via data-flow redirection; includes field multiplication: schematic representation.

Another approach is dummy register usage, in which the first multiplicand is simultaneously stored in a dummy register while being transferred to the multiplier. In the case of a squaring operation, the second multiplicand is then read from this dummy register. Consequently, the registers supplying both multiplicands have different addresses, ensuring that the multiplexer consumes energy even when providing the same value as in the previous cycle, thereby reducing address-based leakage (see Fig. 3). In this case, many but not all gates of the multiplexer are switching, connecting the output of the dummy register to the multiplier, but providing the same value as in the previous clock cycle. Please note that the number of switched gates depends not only on the hamming distance of the addresses of both registers but also on the concrete implementation of the multiplexer, i.e., it is the part of the design which has to be checked carefully.

### III. DISCUSSION AND FUTURE WORK

Each proposed solution has distinct advantages and limitations. The dummy register method for data flow redirection increases design area and costs and slightly raises its energy consumption, but does not influence the execution time of *kP* operations. In contrast, the "re-loading" of the bus solution, which utilizes randomized dummy addressing, requires an extra clock cycle for each instance, thus increasing execution time and energy consumption of the design. Additionally, field multiplication, squaring and addition operations, of different and identical values, can cause the same vulnerability, combining address- and data-bit phenomena. This aspect has to be investigated.

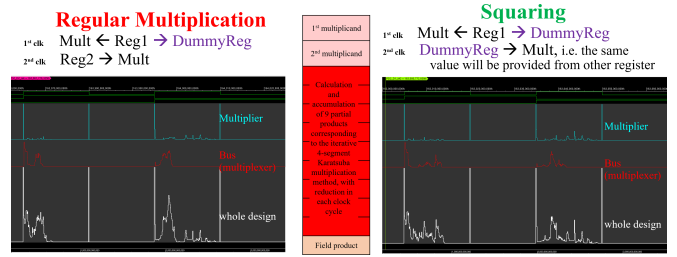

Fig. 3. Mitigating multiplication–squaring distinguishability via bus "re-loading"; includes field multiplication: schematic representation

A promising strategy that can be used in various ways to countermeasure a broad spectrum of physical attacks, including localised fault injection attacks, can be flexible, dynamically reconfigurable redundancy [10].

### ACKNOWLEDGEMENT

This work was funded by EU Project CTIS4NIS, project grant 101249584.